\documentclass[
aip, 
reprint,
floatfix,
natbib]{revtex4-1} %
\usepackage{graphicx}
\usepackage[colorlinks=true, linkcolor=green, citecolor=green]{hyperref}
\usepackage{xcolor}
\usepackage{amsmath}
\usepackage{graphicx}
\usepackage{xcolor}
\usepackage{setspace}
\usepackage{physics}
\usepackage{amsfonts}
\usepackage{float}
\usepackage{tikz}
\usepackage{dirtytalk}
\usetikzlibrary{shapes.geometric, positioning, arrows.meta}
\usepackage{adjustbox}  %
\usepackage[final]{pdfpages}
\usepackage{pgffor}

\renewcommand{\selectlanguage}[1]{}

\makeatletter
\def\fps@figure{t}
\AtBeginDocument{\let\LS@rot\@undefined}
\makeatother

\begin{document}

\title{Harmonic-to-anharmonic thermodynamic integration made simple using REG TI}
\author{Venkat Kapil}%
\email{v.kapil@ucl.ac.uk}

\affiliation{%
    Department of Physics and Astronomy, University College London, 7-19 Gordon St, London WC1H 0AH, UK
}
\affiliation{%
    Thomas Young Centre and London Centre for Nanotechnology, 
    9 Gordon St, London WC1H 0AH, UK
}

\begin{abstract}
Standard harmonic-to-anharmonic thermodynamic integration (TI) is known to develop a near singularity in the integrand for solids exhibiting diffusive degrees of freedom, such as rotating functional groups or migrating defects.
This pathology results in numerical challenges for estimating absolute free energies within a single thermodynamic cycle.
In this work, we introduce a simple regularization that removes this singularity and yields a well-behaved integrand that can be accurately evaluated on a uniform grid. 
The approach — termed Regularized End-point Gradient (REG) TI — is demonstrated on a model system and on predicting the relative stability of paracetamol polymorphs for which quasi-free methyl rotations lead to a near singularity in standard TI. 
We expect REG TI to simplify anharmonic free energy calculations for solids and to potentially enable their automation.

\end{abstract}
	
\maketitle
	
\newpage

The free energy is a central thermodynamic property that determines all equilibrium observables of a system and its thermodynamic stability. 
While equilibrium properties depend ultimately on differences in free energies, often it is desirable to estimate absolute free energies for quantities such as defect formation energies~\cite{cheng_theoretical_2018}, adsorption free energies~\cite{amsler_anharmonic_2021}, and solubilities~\cite{li_computational_2017}, or calculating phase diagrams~\cite{mcbride_phase_2012,reinhardt_quantum-mechanical_2021,kapil_first-principles_2022}.
Unlike differences, absolute free energies can not be estimated from an ergodic atomistic simulation as an ensemble average. 
Instead, they are estimated either as a differences relative to a reference state with known free energy using techniques such as thermodynamic integration (TI)~\cite{kirkwood_statistical_1935}, free energy perturbation~\cite{zwanzig_hightemperature_1954}, or their derivatives~\cite{bennett_efficient_1976, jarzynski_targeted_2002, shirts_statistically_2008}, or by directly evaluating the partition function with more computationally demanding global sampling techniques such as Wang-Landau~\cite{wang_efficient_2001} or nested sampling ~\cite{ashton_nested_2022}. \\

Unfortunately, practical evaluations of absolute free energies are often tedious and computationally demanding, even for solids where a harmonic potential energy surface provides a well-defined reference state.
Consider the TI scheme proposed by ~\citet{frenkel_new_1984}~ -- arguably the most widely used method -- which estimates the anharmonic free energy as the reversible work required to transform the potential from a harmonic reference to the physical one.
This approach yields ill-behaved integrands when the physical configurational ensemble deviates significantly from that of the harmonic reference~\cite{rossi_anharmonic_2016,cheng_computing_2018,kapil_assessment_2019}.
Statistical convergence can be improved, when anharmonic fluctuations are along normal modes, using a harmonic reference obtained from the second-order Taylor expansion of the potential energy surface around a local minimum position~\cite{habershon_free_2011}.
However, many systems involve curvilinear anharmonic motion that is locally perturbed along specific normal modes but larger amplitudes fluctuate into perpendicular directions along other modes~\cite{kapil_assessment_2019}. 
Another extreme case of anharmonic motion is diffusion of atoms in a solid~\cite{cheng_computing_2018}.
These systems exhibit near-singular behavior in the thermodynamic integration integrand at one of the end points, making it challenging to estimate the anharmonic free energy.
One way to estimate the integral, as proposed in the \textit{tour de force} study by ~\citet{rossi_anharmonic_2016}~, is to extrapolate the integrand near its end point by fitting a piecewise rational function over a non-uniform grid with adaptive resolution that is finer close to the end point.
The limitation is that the integrand is prone to systematic errors from the fit, and no general variable transformation exists for constructing an optimal non-uniform grid.
In a later work, ~\citet{cheng_computing_2018} proposed to circumvent this issue by introducing a two-step TI scheme. 
Their approach performs a harmonic-to-anharmonic TI at a low temperature -- where non-ergodic sampling restricts access to large-amplitude curvilinear motion -- and subsequently a low-to-high-temperature TI with an efficient variable transformation and low-variance estimators~\cite{moustafa_very_2015}.
Unfortunately, this procedure requires \textit{a priori} knowledge of the system’s configurational or conformational entropy to correct for the loss of ergodicity at low temperatures~\cite{kapil_assessment_2019}, thereby limiting applicability to generic systems. \\

In this work, we propose a simple and robust approach applicable to complex solids to regularize the integrand for harmonic-to-anharmonic TI.
The proposed approach -- termed  Regularized End-point Gradient (REG) TI -- improves upon existing methods by not requiring multiple TI routes and enables single-shot integration of a smoothly varying integrand. REG TI does not  require \text{a priori} knowledge of the system’s configurational or conformational entropy, a non-uniform grid, or careful fitting of the integrand to a rational approximation.
The origin of the near singularity is first diagnosed on a simple model system and an end-point regularization scheme is introduced to address it.
The approach is then demonstrated on the challenging case of estimating the anharmonic free energies of paracetamol polymorphs exhibiting conformational flexibility~\cite{rossi_anharmonic_2016}.
Finally, we discuss how this approach extends to modern workflows involving machine learning interatomic potentials~\cite{jacobs_practical_2025}, automation~\cite{vandenhaute_machine_2023}, and GPU-accelerated simulations~\cite{cohen_torchsim_2025}. \\

\begin{figure*}
    \centering
    \includegraphics[width=\linewidth]{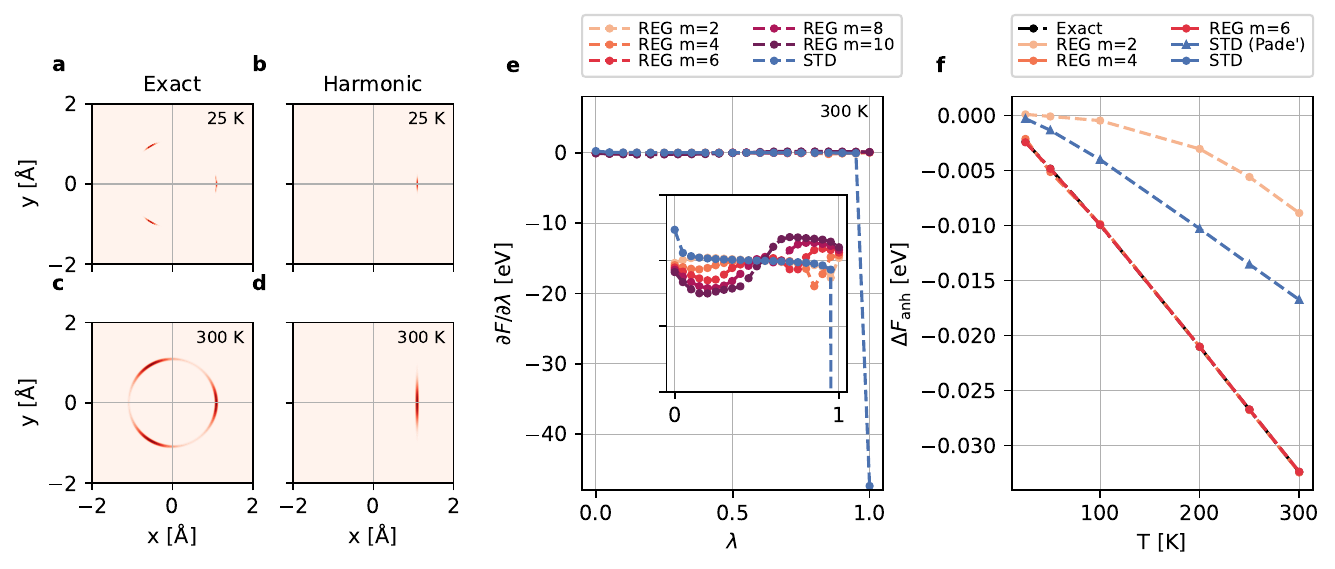}
    \caption{\textbf{Free-energy analysis for the two-dimensional model system of methyl rotation from Ref.~\citenum{kapil_assessment_2019}.}
    Panels \textbf{a}–\textbf{d} show the spatial probability densities at 25\,K and 300\,K for the physical potential and its harmonic reference, with darker colors indicating higher probability density.
    Panel \textbf{e} shows the integrand for standard thermodynamic integration (blue) and for the regularized variant with $m=2,4,6,8,10$ at 300\,K, where darker colors correspond to larger values of $m$.
    The inset magnifies the y-axis to compare the integrand of REG TI for various $m$ with that of standard TI.
    Panel \textbf{f} shows the temperature-dependent anharmonic free energies estimated from standard TI (blue) and its regularized variant for $m=2,4,6$.
    Round markers indicate numerical integration using the trapezoidal rule applied to Eq.~\ref{eq:H_alchemical_eval}, while triangular markers indicate integration via Padé interpolation as detailed in Ref.~\citenum{rossi_anharmonic_2016}.
    The exact anharmonic free energies obtained by direct numerical evaluation of the partition function are shown in black.
    Dashed lines are guide for eyes.}
    \label{fig:cosine_free_energy}
\end{figure*}

We begin by considering a solid described by a classical Hamiltonian with potential energy U($\mathbf{q}$) where
$\mathbf{q}$ is the position vector.
The dependence of the potential on the cell tensor is omitted for brevity.
The classical Helmholtz free energy of the system can be written as
\begin{equation}
    F = F_0 + \Delta_{\text{anh}} F
    \label{eq:deltaF}
\end{equation}
with $F_0$ the classical free energy of a harmonic reference with the potential $U_0(\mathbf{q})$ and $\Delta_{\text{anh}} F$ is the implicitly defined anharmonic free energy correction. 
The dependence of the free energy on thermodynamic variables is omitted for brevity. 
One choice for the harmonic reference, proposed by ~\citet{habershon_free_2011} , is 
\begin{equation}
U_{0}(\mathbf{q}) = U(\mathbf{q}_0) + \frac{1}{2} 
(\mathbf{q} - \mathbf{q}_{0})\cdot
\left.\frac{\partial^{2} U(\mathbf{q})}{\partial \mathbf{q}^{2}}\right|_{\mathbf{q}_{0}}\cdot
(\mathbf{q} - \mathbf{q}_{0})
\end{equation}
with $\mathbf{q}_{0} = \operatorname*{arg\,min}_{\mathbf{q}} U(\mathbf{q})$ denoting the local minimum of the potential.
Within the standard TI approach~\cite{frenkel_new_1984}, the anharmonic free energy correction is estimated as
\begin{equation}
    \Delta F_{\text{anh}} = \int_{0}^{1} \mathrm{d}\lambda~ \left\langle \frac{\partial U(\lambda)}{\partial \lambda}\right\rangle_{\lambda},
    \label{eq:H_alchemical}
\end{equation}
where $\lambda$,  the so-called Kirkwood coupling parameter, defines a fictitious linearly coupled Hamiltonian with the potential
\begin{equation}
    U(\lambda) = \lambda~U + (1 - \lambda)~U_{0},
    \label{eq:linear_coupling}
\end{equation}
where the position dependence of the potentials is omitted for brevity, and $\langle \cdot\rangle_{\lambda}$ denotes a canonical ensemble average for the Hamiltonian with potential $U(\lambda)$.
Eq.~\ref{eq:linear_coupling} allows one to adiabatically transform the harmonic reference into the physical potential, while Eq.~\ref{eq:H_alchemical} expresses the free energy difference as the reversible work required to perform this transformation.
Evaluating Eq.~\ref{eq:H_alchemical} equation for the linearly coupled potential yields
\begin{equation}
    \Delta F_{\text{anh}} = \int_{0}^{1} \mathrm{d}\lambda~ \left\langle U - U_{0}\right\rangle_{\lambda}.
    \label{eq:H_alchemical_eval}
\end{equation}
Eqs.~\ref{eq:deltaF}-\ref{eq:H_alchemical_eval} suggest a straightforward procedure for standard TI: select a set of $\lambda$ values over a grid, perform canonical ensemble simulations with the fictitious Hamiltonian for each $\lambda$, and estimate the integrand as the ensemble average of the difference between the physical and harmonic potentials obtained from these simulations.
The integrand is numerically integrated to obtain the anharmonic free energy correction. \\

This setup unfortunately yields ill-behaved integrands for systems exhibiting curvilinear~\cite{rossi_anharmonic_2016} or diffusive anharmonic motion~\cite{cheng_computing_2018}. 
We illustrate this pathology on a two-dimensional potential
\begin{equation}
    U(r, \theta) = \frac{k}{2}\left(r - r_0\right)^{2} + U_{\theta} \left(1 - \cos{3 \theta}\right),
\end{equation}
a model system defined in Ref.~\citenum{kapil_assessment_2019} for methyl rotation in a paracetamol molecular crystal.
We perform the statistical mechanics of this system numerically on a two-dimensional grid by discretizing $x = r \cos{\theta}$ and $y = r \sin{\theta}$ from -2 to 2 into 1000 bins along each direction. 
We perform TI numerically as well by employing a uniform grid for $\lambda \in [0,1]$ with 20 intervals and the aforementioned configurational space grid.
As can be gathered from  Fig.~\ref{fig:cosine_free_energy}(a), this potential features three local minima that reflect \(C_{3v}\) symmetry associated with the methyl group and thus results in a tri-modal Boltzmann probability distribution. 
In contrast, its harmonic reference features only a single minimum by design, as shown in Fig.~\ref{fig:cosine_free_energy}(b). \\

As temperature is increased, the exact distribution delocalizes in a curvilinear manner (see Fig.~\ref{fig:cosine_free_energy}(c)), while the harmonic distribution delocalizes linearly along the normal modes (see Fig.~\ref{fig:cosine_free_energy}(d)). 
When standard TI is employed, the disparities in the spatial distributions result in a near singularity in the integrand at $\lambda=1$.
This is shown in Fig.~\ref{fig:cosine_free_energy}(e), for the case of 300\,K, while noting that it persists at lower temperatures as well.
The explanation is straightforward: when $\lambda=1$, the integrand $U - U_{0}$ is calculated in the thermodynamic ensemble of the physical potential which samples three local minima.
Configurations drawn from the minima that do not overlap with the harmonic reference yield spuriously large harmonic potential energies and thus the observed near singularity in the integrand. \\

Having established that the pathology originates from the harmonic potential admitting unphysical values at $\lambda=1$, rather than from the configurations being unphysical, we propose a simple remedy to regularize the integrand at the end point.
To achieve this, we first write a general expression of a potential that transforms non-linearly the harmonic potential into the physical one
\begin{equation}
    U(\lambda) = f(\lambda)\,U + g(\lambda)\,U_{0},
    \label{eq:nonlinear_coupling}
\end{equation}
such that $f(0) = 0$, $f(1) = 1$, $g(0) = 1$, and $g(1) = 0$.
While often, for simplicity $f(\lambda)= 1 - g(\lambda)= \lambda$ is considered yielding standard TI, we consider non-linear functions that satisfy an additional boundary condition on the derivative: $\left.\mathrm{d}g(\lambda)/\mathrm{d}\lambda = 0\right|_{\lambda=1}$.
For these functions the integrand, $f'(\lambda)\,U + g'(\lambda)\,U_{0}$, at $\lambda=1$ is simply $f'(\lambda) U$ as opposed to $U - U_0$ for standard TI.
Thus, the spuriously high contributions from $U_{0}$ are suppressed in the integrand through its vanishing coefficient.
We note that the proposed regularization is related to an approache used in adiabatic free-energy dynamics, where the same pathology emerges as a large barrier in $\lambda$-space~\cite{abrams_efficient_2006}; we became aware of this work only after preparing the manuscript. \\

In this work, we explore a simple family of non-linear functions: $f(\lambda) = \lambda^{m}$ and $g(\lambda) = (1 - \lambda)^{m}$, where $m$ is an integer greater than one, which yields
\begin{equation}
    \Delta F_{\text{anh}}  = \int_{0}^{1} \mathrm{d}\lambda~ m\,\left\langle \lambda^{m-1} U - (1 - \lambda)^{m-1} U_{0} \right\rangle_{\lambda}.
    \label{eq:H_alchemical_eval_reg}
\end{equation}
Here, $\langle \cdot \rangle_{\lambda}$ denotes a canonical ensemble average for the Hamiltonian implicitly defined by Eq.~\ref{eq:nonlinear_coupling}.
As in standard TI, $\lambda = 1$ implies $g(1) = 0$, meaning that the integrand is evaluated in the physical ensemble. However, the integrand $m~U$ is no longer dependent on $U_{0}$, which in the standard approach admits spuriously high values.
While introducing a non-linear functional form for $f$ is not required to address the said pathology, enforcing $\left.\mathrm{d}f(\lambda)/\mathrm{d}\lambda\right|_{\lambda=0} = 0$ provides a minor computational advantage.
The integrand at $\lambda=0$ is $m~U_0$, which does not depend on the physical potential and thus avoids high variance from evaluating potential energies for physically distorted configurations displaced linearly along the normal modes~\cite{kapil_assessment_2019}.
We refer to this approach as Regularized End point Gradient (REG) TI.
In Appendix~\ref{appendix}, we show analytically, for an extreme case of a harmonic-to-ideal-gas TI in one dimension, that REG TI eliminates the near singularity for $m > 1$. \\

As shown in Fig.~\ref{fig:cosine_free_energy}(e), REG TI eliminates the near-singularity for all considered values of $m$.
For $m=2$, the integrand exhibits a small magnitude at $\lambda=1$, however, as seen in the inset, it exhibits a sharp variation close to $\lambda=1$ due to the finite contribution of $U_{0}$ from configurations sampled in the other minima.
This sharpness is attenuated by increasing $m$.
Here, we consider even values of $m$ up to 10 and observe a smoothly varying integrand for $m=6,8,10$, with the magnitude of the integrand increasing with $m$.
We further examine the impact of the integrand on the computed anharmonic free-energy difference in Fig.~\ref{fig:cosine_free_energy}(f) as a function of temperature.
Naive trapezoidal integration for standard TI yields unphysical values outside the $y$-range.
We also consider Padé interpolation as proposed in Ref.~\citenum{rossi_anharmonic_2016} by fitting a piece-wise rational function to the integrand, extrapolating end-point values and performing numerical integration of the piece-wise function.
While this procedure improves the results compared to trapezoidal integration, it is noted to systematically underestimate the free energy and the conformational entropy (the reduced magnitude of the slope).
In contrast, the REG TI yields a well-behaved integrand that can be directly integrated using the trapezoidal rule, achieving quantitative agreement with the reference for $m \ge 4$ up the largest value considered.
We do not consider $m>10$ due to a practical trade-off. 
Larger $m$ values more strongly suppress the contribution of $U_0$ near $\lambda = 1$, but also amplify the derivative of the coupling, which can induce oscillatory behavior and thus require a finer quadrature to accurately estimate the integrand. \\

\begin{figure}
    \centering
    \includegraphics[width=0.5\textwidth]{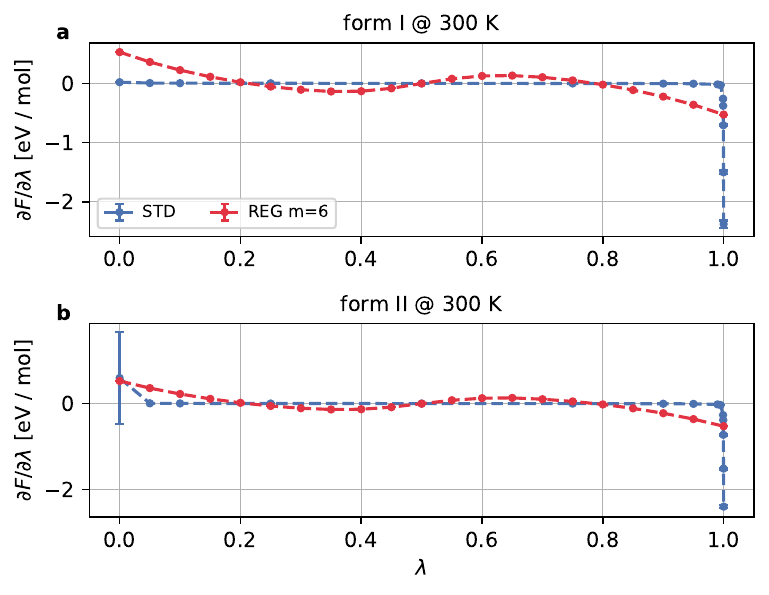}
    \caption{\textbf{Standard and regularized thermodynamic integration integrands for paracetamol polymorphs.} Panels \textbf{a} and \textbf{b} show the integrands corresponding to standard (blue) and regularized thermodynamic integration with $m=6$ (red) at 300\,K, using the experimental lattice parameters for forms~I and~II, respectively. Error bars indicate a 2$\sigma$ confidence interval estimated via block averaging using five blocks. Dashed lines are guide for eyes.}
    \label{fig:paracetamol_free_energy}
\end{figure}

As a demonstration on a realistic system, we apply REG TI to estimate the anharmonic free energies of two polymorphs of paracetamol.
This system is particularly interesting as it exhibits conformational flexibility -- an aspect that poses a challenge for accurately evaluating anharmonic free energies in pharmaceutically relevant crystalline materials -- in the form of methyl rotation, and has been examined in previous thermodynamic integration studies as a challenging system~\cite{rossi_anharmonic_2016, kapil_assessment_2019}.
We adopt the simulation setup from Ref.~\citenum{rossi_anharmonic_2016}, where Form~I of paracetamol is monoclinic and contains four molecules per unit cell, while Form~II is orthorhombic with eight molecules per unit cell. The potential energy surface is modeled using the Merck molecular force field (MMFF) implemented in LAMMPS~\cite{thompson_lammps_2022}.
REG TI is performed on a uniform grid of 20 $\lambda$ values with $m$=6, while standard TI is performed on a non-uniform grid proposed by the authors in Ref.~\citenum{rossi_anharmonic_2016}.
Each value of $\lambda$ is treated by a distinct molecular dynamics simulation in the canonical ensemble at 300\,K and the experimental volume. 
The i-PI code~\cite{litman_i-pi_2024} is used to perform the simulations with a weighted sum of the harmonic and physical potentials with with weighting being different for standard and REG TI. 
Molecular dynamics simulations are run for 1\,ns with a 0.5\,fs time step, and the harmonic and physical potentials are sampled every 20 steps.
An optimal sampling generalized Langevin equation thermostat~\cite{ceriotti_colored-noise_2010} is employed to enforce efficient sampling in the canonical ensemble. \\

Fig.~\ref{fig:paracetamol_free_energy} shows the integrands for the two polymorphs. 
The integrand for standard TI exhibits a near-singularity at $\lambda = 1$ for both forms and a high variance at $\lambda = 0$ for Form~II -- arising from the physical potential being evaluated on outlier configurations with harmonic fluctuations orthogonal to the direction of methyl rotation.
The REG TI yields a well-behaved integrand, with the pathological behavior at both ends eliminated, allowing the anharmonic free-energy corrections to be estimated for each phase using the trapezoidal rule.
We obtain $\Delta_{\text{anh}}[\mathrm{I}] = -42.8 \pm 4.9$\,meV/molecule $\equiv -4.1 \pm 0.5$\,kJ/mol and $\Delta_{\text{anh}}[\mathrm{II}] = -48.2 \pm 2.4$\,meV/molecule $\equiv -4.6 \pm 0.2$\,kcal/mol, both in excess of 1\,kcal/mol.
Here, the uncertainties correspond to a 2$\sigma$ confidence interval estimated via block averaging using five blocks.
We next quantify the extent to which classical anharmonic motion alters the relative stability of form~I with respect to form~II: $\Delta\Delta_{\text{anh}} = \Delta_{\text{anh}}[\mathrm{I}] - \Delta_{\text{anh}}[\mathrm{II}] = 5.4 \pm 5.5$\,meV/molecule $\equiv 0.5 \pm 0.5$\,kJ/mol.
This spectacular cancellation of anharmonic contributions to the relative stability is a consequence of the simple harmonic intramolecular bond and angle terms employed in MMFF.
In contrast, first-principles (quality) potential energy surface can yield net anharmonic free-energy contributions~\cite{rossi_anharmonic_2016, hoja_reliable_2019, pia_accurate_2025} as the implicit intramolecular energy components are sensitive to intermolecular interactions and do not systematically cancel across polymorphs. \\

It is instructive to compare our results with earlier work on paracetamol polymorphs employing MMFF.
Ref.~\citenum{rossi_anharmonic_2016} reports anharmonic corrections relative to the gas phase, making it feasible to compare only $\Delta\Delta_{\text{anh}}$.
We recover this value by subtracting the reported free energies of form~I relative to form~II from Table~I, estimated classically at harmonic and anharmonic levels.
We find $\Delta\Delta_{\text{anh}} = 1.0 \pm 0.7$\,meV/molecule $\equiv 0.1 \pm 0.1$\,kJ/mol, where the uncertainty is estimated by assuming an error of 0.5\,meV, as the reported values are rounded to the nearest meV.
We consider this agreement to be fair, noting that discrepancies exceeding 10\,meV/molecule were reported in Ref.~\citenum{kapil_assessment_2019} using a setup that employed an alternative TI route, slightly different equilibrium volumes for the polymorphs, and a finer PPPM mesh~\cite{hockney_computer_2021}. \\

To summarize, we present a single-shot harmonic-to-anharmonic TI scheme -- referred to as REG TI -- which yields well-behaved integrands for solids exhibiting diffusive or curvilinear anharmonic motion.
We expect this approach to be advantageous over existing approaches. 
Compared to multi-step schemes,  which perform harmonic-to-anharmonic TI at low temperatures followed by TI over temperature, our method does not require \textit{a priori} knowledge of the configurational or conformational or conformational entropy and is not prone to compounded errors across multiple TIs.
Furthermore, it requires no additional setup or analysis overhead over standard TI extending it's simplicity to strongly anharmonic systems.
Another appealing prospect is its direct applicability to quantum harmonic-to-anharmonic free-energy calculations within the path-integral framework, as demonstrated in Ref.~\citenum{kapil_complete_2022} for rigid solids. 
The REG TI can enable single-shot evaluation of quantum anharmonic free energies for complex systems, avoiding a separate classical-to-quantum TI as is usually done~\cite{fang_inverse_2016,habershon_thermodynamic_2011}. \\

A limitation of this work is that the chosen switching function does not flatten the integrand into a straight line, which would allow accurate evaluation of the integrand with fewer $\lambda$ points. 
Thus, the approach has not yet been fully optimized for efficiency, and there may exist switching functions that further reduce the computational cost of the TI further.
However, even in its current form, it offers improved reliability and numerical stability compared to existing methods.
Moreover, as in standard TI, the calculations can be trivially parallelized as independent CPU runs or efficiently batched on GPUs, as is now possible for molecular dynamics with machine-learning interatomic potentials~\cite{cohen_torchsim_2025}. 
Together with recently developed automation frameworks for free-energy calculations that incorporate active learning for training machine-learning interatomic potentials~\cite{vandenhaute_machine_2023}, these features make REG TI a robust foundation for automated and high-throughput first-principles anharmonic free-energy calculations with broad applicability. \\

\section*{Acknowledgments}
We thank Michele Ceriotti, Jutta Rogal, Diptarka Hait, Sander Vandenhaute, Daan Frenkel, Mark Tuckerman, and Benjmain Xu Shi, for valuable comments on the manuscript and insightful discussions. Notably, we thank Mark Tuckerman for making us aware of their earlier work on adiabatic free-energy dynamics, where a similar regularization was used to democratically sample alchemical space. VK acknowledges the hospitality of the Initiative for Computational Catalysis at the Flatiron Institute during a visit, where a portion of this research was carried out. \\

\subsection*{Conflict of Interest}
There are none to declare. \\

\section*{Data availability}
Scripts to setup REG TI are available at \href{https://github.com/venkatkapil24/data-REG TI}{https://github.com/venkatkapil24/data-REG TI}. \\

\appendix
\section{harmonic to ideal gas TI}
{\label{appendix}}

We apply TI while transforming a harmonic potential $U_{0}(x) = \frac{1}{2} k x^{2}$ within a finite configurational space $x \in [-a, a]$ into a fully diffusive ideal gas $U_{0}(x) = 0$ in the same configurational space. 
We show analytically that REG TI removes the near-singularity that arises at the end point in standard TI.

We first define a fictitious potential $
U(x, \lambda, m) = \tfrac{1}{2}\,k\,(1-\lambda)^m\,x^2,$
for an integer $m \ge 1$, that transforms the system form a harmonic oscillator for $\lambda=0$ to an ideal gas for $\lambda=1$.
Standard TI corresponds to the case of $m=1$ and REG TI corresponds to $m > 1$. The TI integrands can be expressed as
\begin{equation}
I(\lambda, m) = -\frac{m k}{2}(1-\lambda)^{m-1}\,\langle x^2 \rangle_{\lambda, m} \nonumber
\label{eq:I_general}
\end{equation}
with $\left<\cdot\right>_{\lambda, m}$ corresponds to a canonical ensemble average for the potential $U(x, \lambda, m) $. Since the configurational space domain is bound in $x \in [-a, a]$, the expectation value of $x^2$ at inverse temperature $\beta$ is estimated as a Gaussian integral with finite limits
\begin{widetext}
\begin{equation}
\langle x^2\rangle_\lambda =
\frac{1}{\beta k (1-\lambda)^m}
-\frac{a\,\exp\left(-\tfrac{1}{2}\beta k (1-\lambda)^m a^2\right)}
{\sqrt{\pi}\,\sqrt{\tfrac{\beta k}{2}(1-\lambda)^m}\;
\mathrm{erf}\!\left(a\sqrt{\tfrac{\beta k}{2}(1-\lambda)^m}\right)},
\label{eq:x2_mean} \nonumber
\end{equation}
\end{widetext}
with $\mathrm{erf}\left({x}\right)$ the Gauss error function. 

To understand if the integrand exhibits a near singularity at $\lambda=1$, it is instructive to estimate the ratio $R(m) = {I(1,m)}~/~{I(0, m)}$. For a generic harmonic constant and box size, the integrand at $\lambda=0$ is 
\begin{equation}
I(0,m)
= -\frac{m}{2\beta}
+ \frac{m k a}{2}
\frac{\exp\!\left(-\tfrac{1}{2}\beta k a^{2}\right)}
{\sqrt{\pi}\,\sqrt{\tfrac{\beta k}{2}}\,
\mathrm{erf}\!\left(a\sqrt{\tfrac{\beta k}{2}}\right)}\,. \nonumber
\end{equation}
In the regimes of large $k$ (stiff force constant) or large $a$ (large box), the exponent in the second term approaches zero while the error function approaches unity, yielding
\begin{equation}
    I(0,m) \approx -\frac{m}{2\beta}. \nonumber
\end{equation}

To estimate the integrand at $\lambda=1$, we first define $\epsilon = (1-\lambda)$. The integrand for generic $\epsilon$ is
\[
    I(1-\epsilon, m) = -\frac{m k}{2}
\left[
\frac{1}{\beta k\,\varepsilon}
-\frac{a\,\varepsilon^{\tfrac{m}{2}-1}\exp\!\left(-\tfrac{1}{2}\beta k a^{2}\varepsilon^{m}\right)}
{\sqrt{\pi}\,\sqrt{\tfrac{\beta k}{2}}\,
\mathrm{erf}\!\left(a\sqrt{\tfrac{\beta k}{2}}\,\varepsilon^{m/2}\right)}
\right]
\]
For non-vanishing $k$ and $a$ and in the limit $\varepsilon \to 0^{+}$, the exponent in the numerator of the second term and the inverse of the error function in the denominator of the second term can be expanded up to $\varepsilon^{2m}$ as
\begin{align}
\exp\!\left(-\tfrac{1}{2}\beta k a^{2}\varepsilon^{m}\right)
& = 1 - \tfrac{1}{2}\beta k a^{2}\varepsilon^{m} + O(\varepsilon^{2m}), \nonumber \\  
\frac{1}{\mathrm{erf}\!\left(a\sqrt{\tfrac{\beta k}{2}}\,\varepsilon^{m/2}\right)}
& = \frac{\sqrt{\pi}}{2\,a\sqrt{\tfrac{\beta k}{2}}\,\varepsilon^{m/2}}
\left[
1 + \right. \nonumber  \\ 
& \left.\frac{1}{3}\left(a^{2}\tfrac{\beta k}{2}\right)\varepsilon^{m}
+ O(\varepsilon^{2m})
\right]. \nonumber
\end{align}
Collecting terms up to $O\!\left(\varepsilon^{2m-1}\right)$, the second term in the integrand reduces to 
\begin{equation*}
\frac{a\,\varepsilon^{\tfrac{m}{2}-1}\,
\exp\!\left(-\tfrac{1}{2}\beta k a^{2}\varepsilon^{m}\right)}
{\sqrt{\pi}\,\sqrt{\tfrac{\beta k}{2}}\,
\mathrm{erf}\!\left(a\sqrt{\tfrac{\beta k}{2}}\,\varepsilon^{m/2}\right)}
= \frac{1}{\beta k\,\varepsilon}
\left[
1 - \frac{1}{3}\,\beta k a^{2}\,\varepsilon^{m}
\right],
\end{equation*}
leading to the simplified expression
\begin{equation*}
I(1-\varepsilon, m) = -\frac{m k a^{2}}{6}\,\varepsilon^{\,m-1}
+ O\!\left(\varepsilon^{\,2m-1}\right).
\end{equation*}
When $m=1$, the leading-order term is independent of $\varepsilon$, and thus $I(1,1) = -k a^{2}/6$. For $m>1$ the integrand is suppressed with 
$I(1,m) = \lim_{\varepsilon \to 0^{+}} I(1-\varepsilon,m) = 0$.

We can now estimate the ratio $R(m) = I(1,m) / I(0,m)$ for standard ($m = 1$) and REG TI ($m > 1$).
In the regime of large $k$ or $a$, the ratios become
\[
R(1) \approx \frac{\beta k a^{2}}{3},
\qquad
R(m>1) \approx 0,
\]
indicating that for stiff force constants or large box sizes the standard TI integrand exhibits a near singularity, whereas the REG TI formulation removes this near singular behavior.

\section*{References}

\end{document}